\documentclass[aps,pre,twocolumn,reprint,groupedaddress]{revtex4-1}

\usepackage{amsmath,graphicx,amssymb}
\usepackage{url}
\usepackage{verbatim}
\usepackage{float}
\usepackage{epstopdf}
\usepackage{amsfonts}
\usepackage{amssymb,bm}
\usepackage{caption}
\usepackage{subcaption}
\usepackage{color}

\newcommand{\beq}{\begin{equation}}
\newcommand{\eeq}{\end{equation}}

\newcommand{\vu}{\vec{u}}
\newcommand{\vb}{\vec{b}}
\newcommand{\vk}{\vec{k}}

\newcommand{\vf}{\vec{f}}

\newcommand{\epk}{\varepsilon_K}
\newcommand{\epm}{\varepsilon_M}
\let\oldhat\hat
\renewcommand{\vec}[1]{{\bm{#1}}}
\renewcommand{\hat}[1]{{\oldhat{\bm{#1}}}}

\begin{document}

\title{Fully-resolved array of simulations
investigating the influence of the magnetic Prandtl number
on MHD turbulence}


\author{Mairi E. McKay}
\email[]{mairi.mckay@ed.ac.uk}
\author{Arjun Berera}
\email[]{ab@ph.ed.ac.uk}
\author{Richard D.J.G. Ho} 
\email[]{richard.ho@ed.ac.uk}
\affiliation{School of Physics and Astronomy, University of Edinburgh, 
James Clerk Maxwell Building, Peter Guthrie Tait Road, 
Edinburgh, EH9 3FD, Scotland}

\date{\today}

\begin{abstract}
We explore the effect of the magnetic Prandtl number Pm on energy and dissipation
in fully-resolved direct numerical simulations of 
steady-state, mechanically-forced homogeneous magnetohydrodynamic turbulence
in the range $1/32<$Pm$<32$.
We compare the spectra and show that if the simulations are not fully resolved,
the steepness of the scaling of the kinetic-to-magnetic dissipation ratio with Pm
is overestimated.
We also present results of decaying turbulence with helical and nonhelical magnetic fields,
where we find nonhelical reverse spectral transfer for Pm$<1$ for the first time.
The results of this systematic analysis have applications ranging from
stars, planetary dynamos, and accretion disks.
\end{abstract}

\pacs{47.65.-d,52.30.Cv,47.27.Gs,47.27.ek} 

\maketitle

\section{Introduction}

Turbulence is observed in an enormous variety of situations
but fully understood in few.
When an electrically-conducting fluid is exposed to a magnetic field,
the turbulent dynamics can be described by the magnetohydrodynamic (MHD) equations,
which dictate how the two main aspects of the fluid
(the velocity and magnetic fields) interact.
The seminal work on MHD was done by Hannes Alfv\'en \cite{Alfven1942},
earning him the Nobel Prize.
MHD offers valuable insights into
astrophysical and geophysical phenomena, including the solar wind
and the Earth's magnetic field, and aids the development of
industrial processes such as fusion reactors \cite{Biskamp2003,Davidson2001,Frisch1995,Verma2004,Keppens2010}.

Physical properties of a magnetofluid affect its behaviour. 
One such property is the magnetic Prandtl number Pm $= \nu/\eta$,
where $\nu$ is the kinematic viscosity and $\eta$ the magnetic resistivity,
which is a material property of the fluid.
We may also write Pm $=\text{Rm}/\text{Re}$,
where Rm and Re are the magnetic and kinetic Reynolds numbers,
quantifying respectively the turbulence of the magnetic and kinetic 
components of the fluid. 
In nature, extreme values of Pm are commonplace: 
stellar and planetary interiors are in the range Pm $\sim 10^{-4}$ to $10^{-7}$
and smaller, while the interstellar medium and 
cosmological-scale magnetic fields have estimated values of Pm $\sim 10^{10}$ to $10^{14}$ 
\cite{Heliophys,Verma2004,Plunian2013,Federrath2014,Schober2012,Schober2018}.
The achievable range of Pm in direct numerical simulations (DNS) is highly restricted 
because of computational requirements and is often set to one,
which is not representative of most magnetofluids. 
Extrapolating from simulations with Pm in the vicinity of one
is often necessary when connecting computational results
to real-life applications.
That said, the region around unity is not without its applications:
black hole accretion disk models indicate
that Pm may transition from being very small in most of the disk,
to being greater than one near the centre,
which may explain the change of state
from emission to accretion in these objects \cite{Balbus2008}.
Estimates of Pm in the solar wind and solar convective zone are Pm $\simeq 1$ \cite{Verma1996,Verma2004}.

In this paper we present an array of 36 high-resolution DNS
of mechanically-forced, homogeneous, incompressible magnetohydrodynamic turbulence
without a mean magnetic field, with $1/32<$Pm$<32$.
Additionally, we present 18 decaying simulations with $1/16<$ Pm $<16$, 
in which we test the effect of Pm on
reverse spectral energy transfer 
(which includes any transfer of energy from small to large scales is
not restricted to just inverse cascade).
With our forced data we focus on the
energy spectra, 
the ratios of the total kinetic and magnetic energies $E_K/E_M$,
called the Alfv\'en ratio,
and
the kinetic and magnetic dissipation rates $\epk/\epm$.
We also discuss resolution requirements in connection with 
recent theoretical findings.

In previous studies, an approximate scaling $\epk/\epm\simeq\text{Pm}^q$ was found \cite{Brandenburg2014,Sahoo2011}.
The parameter $q$ varied depending on the magnetic helicity (which includes the knottedness of the magnetic field,
and contributions from twist, writhe, and linkage \cite{Berger1984,Berger1999}) and whether 
Pm was greater than or less than one.
However,
these papers only guaranteed full resolution of one dissipation scale.
In other words, the largest wavenumber in the simulation, $k_{max}$,
was greater than either the kinetic dissipation wavenumber $k_\nu=(\epk/\nu^3)^{1/4}$ 
or the magnetic dissipation wavenumber $k_\eta=(\epm/\eta^3)^{1/4}$,
but not both.
This is an issue because although a
system's energy is mostly concentrated in the largest length scales,
the dissipation spectrum is proportional to the wavenumber squared.
In hydrodynamic turbulence, in order to capture 99.5\% of the dissipative dynamics,
the condition $k_{max}>1.25 k_\nu$ must be fulfilled \cite{Yoffe2012,Donzis2008,Wan2010}.
This was our definition of 'fully-resolved' and in all our forced simulations we had both $1.25k_\nu<k_{max}$
and $1.25k_\eta<k_{max}$.
This paper also gives an explanation for the scaling which has not been done before.


Our set of forced simulations are an extensive dataset for
DNS of homogeneous MHD turbulence, with 36 data points in the Re-Rm plane covering a
square grid (see Fig.~\ref{fig:params}). Re and Rm range from approximately 50 to 2300, 
allowing for a three order of magnitude range in magnetic Prandtl number.
Each point was run on a $512^3$ or $1024^3$ lattice depending on
individual resolution requirements, 
ensuring all data was fully resolved.
This is the largest fully-resolved dataset for a Pm study.

Large values of magnetic helicity
 encourage reverse spectral transfer (RST),
 where energy is transferred to the
largest length scales in the system, 
rather than to the small, dissipative scales, as in the usual Richardson-Kolmogorov phenomenology
\cite{Leorat1975,Pouquet1976,Alexakis2005a,Alexakis2007}.
Whilst RST does not imply an inverse cascade, an inverse cascade is a type of RST.
The second aspect of our study covers magnetofluids with nonzero magnetic helicity. 
We found RST in both helical and nonhelical turbulence down to $\text{Pm}=1/4$,
increasing as Rm increased, with Re playing little role.
We thus confirm the results of recent simulations that found 
RST without helicity \cite{Berera2014,Brandenburg2017,Brandenburg2015}
and have seminal results showing RST occurring for $\text{Pm}<1$.

\section{Simulations}

We carried out DNS of the incompressible MHD equations
\begin{align}
  &\partial_t \vu=-\nabla P - (\vu\cdot\nabla)\vu + (\vb\cdot\nabla)\vb + \nu \nabla^2 \vu + \vf \ ,\\
  &\partial_t \vb=\nabla \times (\vu \times \vb) +\eta\nabla^2\vb \ ,\\
  &\nabla\cdot\vu=0\text{, }  \nabla\cdot\vb=0 \ ,
\end{align}
where $\vu$ is the velocity field, $\vb$ is the magnetic field in Alfv\'{e}n units,
$P$ is the total pressure, the density is constant and set to 1,
and $\vf$ is a random force
defined via a helical basis:
\begin{equation}
\label{eq:AHF}
\vf(\vk,t) =
 A(\vk)\vec{e}_1(\vk,t)+B(\vk)\vec{e}_2(\vk,t),
\end{equation}
where $\vec{e}_1\cdot\vec{e}_2^*=\vec{e}_1\cdot\vk=\vec{e}_2\cdot\vk=0$
and $\vec{e}_1$ and $\vec{e}_2$ are unit vectors statisfying
 $i\vk\times\vec{e}_1=k\vec{e}_1$ and $i\vk\times\vec{e}_2=-k\vec{e}_2$ 
\cite{McKay2017,Waleffe1992,Lesieur1987}.
$A(\vk)$ and $B(\vk)$ are variable parameters that allow the injection of helicity 
to be adjusted; we chose to force nonhelically.
We solved the MHD equations numerically using a pseudospectral, fully-dealiased code
(see \cite{Yoffe2012,Linkmann2016} for details)
on a three-dimensional periodic domain.
The initial fields were random Gaussian with magnetic and kinetic energy spectra of the form
$E_{M,K}(k,t=0)=Ck^4\exp(-k^2/(2k_0)^2)$,
where $C$ is a positive real number and $k_0$ is the peak of the spectrum.
In our forced simulations we set $k_0=5$ and
forced the velocity field at the largest scales, $1\leq k\leq2.5$.
The nature of the forcing function and the forcing length scale 
do not greatly affect the dynamics \cite{McKay2017,Brandenburg2018}.
We also ran decaying simulations, where we were less interested in the inertial range
energy spectra and more interested in RST,
so we set the peak at $k_0=40$.
There was no imposed magnetic guide field.
The viscosity and resistivity of each simulation are given in Fig.~\ref{fig:params};
note that Rm$\simeq 0.65/\eta$ and Re$\simeq 0.65/\nu$.
This value of 0.65 comes from the fact that the rms velocity $u$
and integral length scale $L$ are relatively constant during the simulations,
with Re $= uL/\nu$ and Rm = $uL/\eta$.

\section{Results}

\subsection{Energy}

\begin{figure}
  \includegraphics[width=\linewidth]{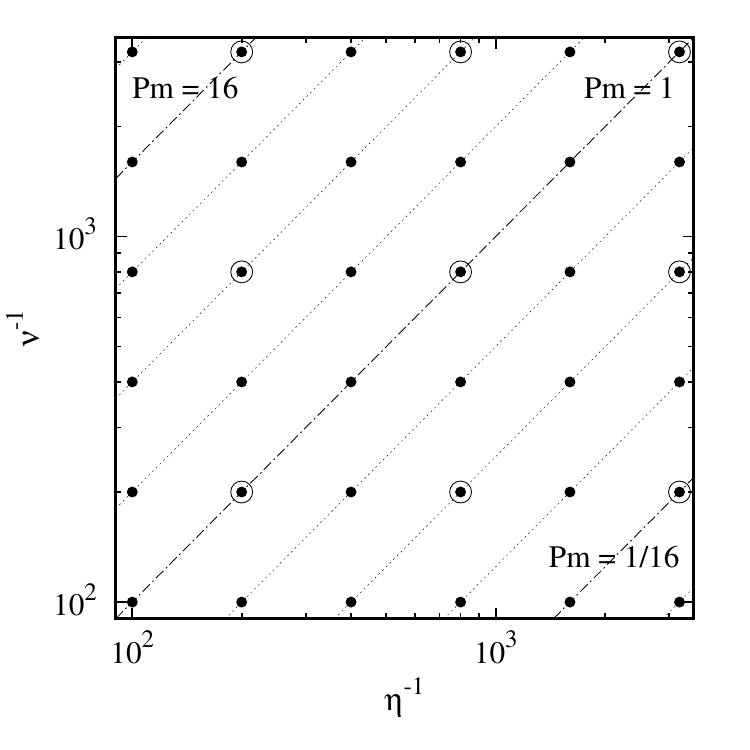}
  \caption{\label{fig:params}Small circles show $\nu^{-1}$ and $\eta^{-1}$ for each of the 36 simulations appearing
  in Figs.~\ref{fig:spec}, \ref{fig:pm_E} and \ref{fig:pm_e}.
  The 9 large circles indicate the decaying helical and nonhelical simulations 
  with initial spectra peaking at $k_0=40$ 
  (see Fig.~\ref{fig:IC}). 
  The lines indicate points of constant $\text{Pm}=2^n$ for $-5\leq n \leq 5$.
  The largest and smallest values of $\eta$ and $\nu$ are 0.01 and 0.0003125.}
\end{figure}
\begin{figure}
  \begin{subfigure}{\linewidth}
     \caption{\label{fig:kin_spec}}
    \includegraphics[width=\linewidth]{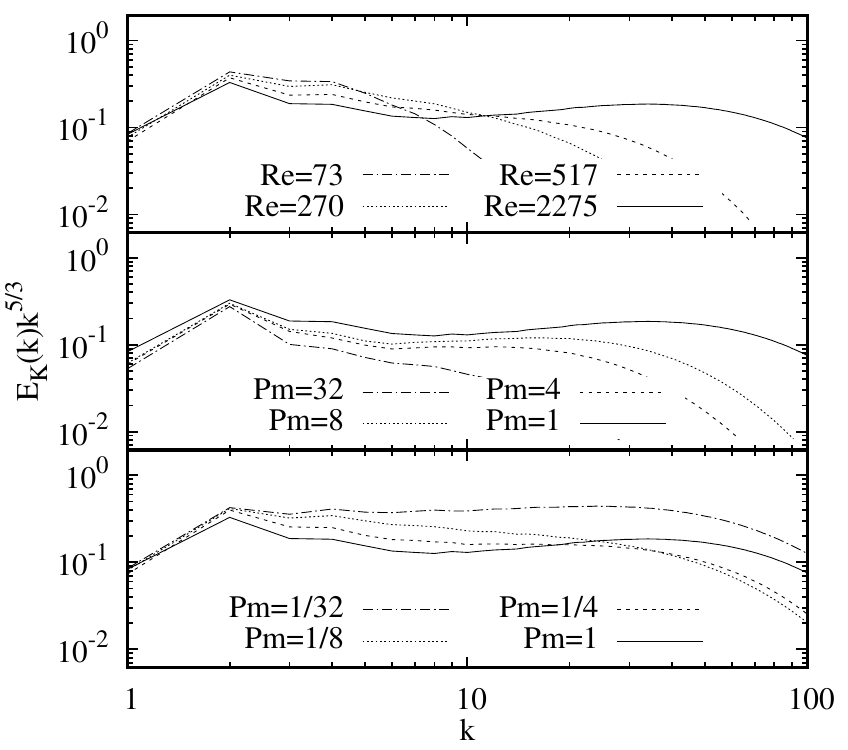}
  \end{subfigure}

  \begin{subfigure}{\linewidth}
    \caption{\label{fig:mag_spec}}
    \includegraphics[width=\linewidth]{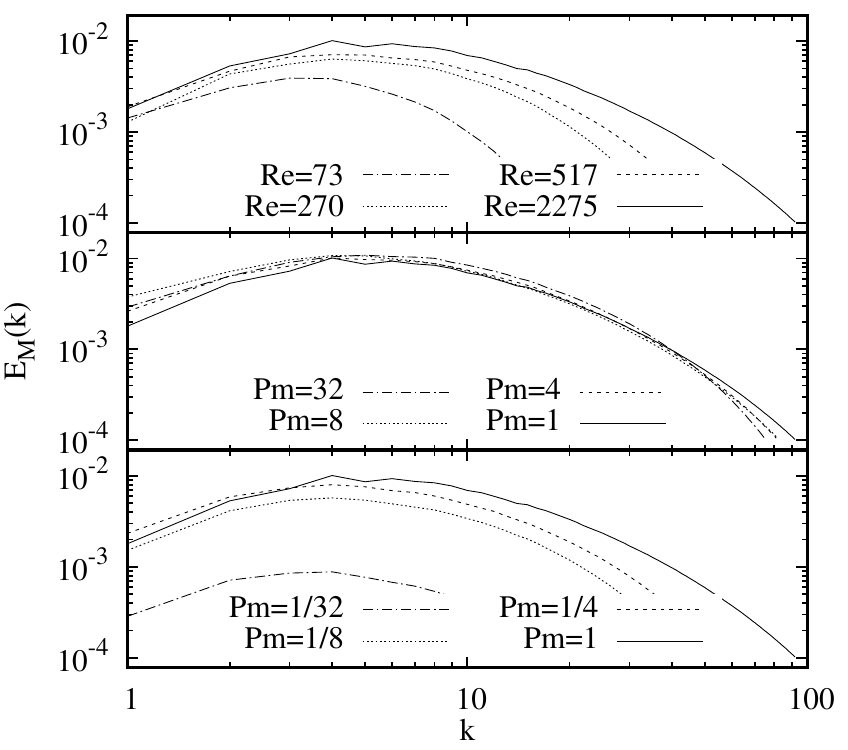}
  \end{subfigure}
  \caption{\label{fig:spec} Selected simulations, (a) shows kinetic energy spectra,
    compensated by $k^{-5/3}$, (b) shows uncompensated magnetic energy spectra.
The top images show data with $\text{Pm}=1$; the second show data with Rm $\simeq$ 2275
  and the third show Re $\simeq$ 2275.
  In each plot the solid line corresponds to the same simulation,
  with $\text{Pm}=1$ and Re=Rm $\simeq$ 2275.}
\end{figure}
Figure \ref{fig:kin_spec} shows the time-averaged compensated
kinetic energy spectra of selected simulations.
In each of the three plots the solid line represents
the same simulation, with Re=Rm$\simeq2275$ and $\text{Pm}=1$.
The top plot shows the spectra of four simulations where Re
and Rm were increased with $\text{Pm}=1$ kept constant.
The middle plot compares data with 
Rm $\simeq$ 2275 and Pm increasing from 1 to 32 by decreasing Re;
 while the bottom plot shows data with Re $\simeq2275$ and
Pm being decreased from 1 to 1/32 via decreasing Rm.
When we increase Re but keep Pm constant, as in the top plot, we see that
less energy is stored in the large scales of the velocity field,
whereas if we increase Re but keep Rm constant and large-valued,
as in the middle plot,
the amount of energy in the large-scale velocity field is slightly enhanced.
The spectrum most closely resembling
the Kolmogorov $k^{-5/3}$ scaling is the $\text{Pm}=1/32$ run
in the bottom plot, which seems to be below the dynamo action onset threshold,
and so the magnetic
field (which was initially in equipartition with the
velocity field) will eventually decay completely, leaving a purely
hydrodynamic simulation.

The corresponding magnetic energy spectra 
are shown in Fig.~\ref{fig:mag_spec}. The spectra are 
most heavily influenced by Rm. 
In the top and bottom plots, Rm is varied while $\text{Pm}$ and Re
are respectively kept constant. The spectra produced 
in these two plots are relatively
similar except in the Rm=73 case, 
where for $\text{Pm}=1$ the magnetic field is sustained
but for $\text{Pm}=1/32$ it is decaying. In the second plot we see that increasing Pm
with constant Rm may slightly augment the large-scale magnetic field. 
Whilst this appears to imply Pm-dependence of the energy spectra,
the total energy spectra $E_T(k)=E_K(k) + E_M(k)$ 
(equivalent to thinking in terms of Els\"asser variables)
appears to depend only on the maximum of Re or Rm, and is thus independent of Pm.

\begin{figure}
  \includegraphics[width=\linewidth]{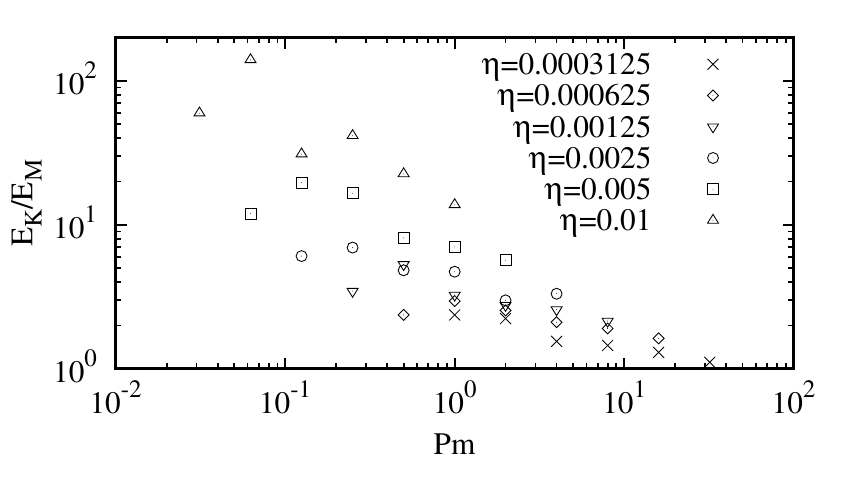}
  \caption{\label{fig:pm_E}Time-averaged 
 Alfv\'en ratios of simulations
  grouped according to resistivity, $\eta$.}
\end{figure}
Figure \ref{fig:pm_E} shows the time-averaged 
Alfv\'en
ratios as a function of Pm, grouped into sets of points with approximately equal Rm.
For fixed Rm the 
Alfv\'en ratios tend to decrease as Pm is increased,
although the slope flattens at larger Rm.
Bearing in mind that Rm doubles with each set of points,
we see that the data are converging onto an asymptotic high-Rm limit.
For all values of Pm,
the ratio $E_K/E_M$ decreases with increasing Rm. These behaviours are in agreement with 
what was put forward in Ref.~\cite{Haugen2003}.

\subsection{Dissipation}

\begin{figure}
  \includegraphics[width=\linewidth]{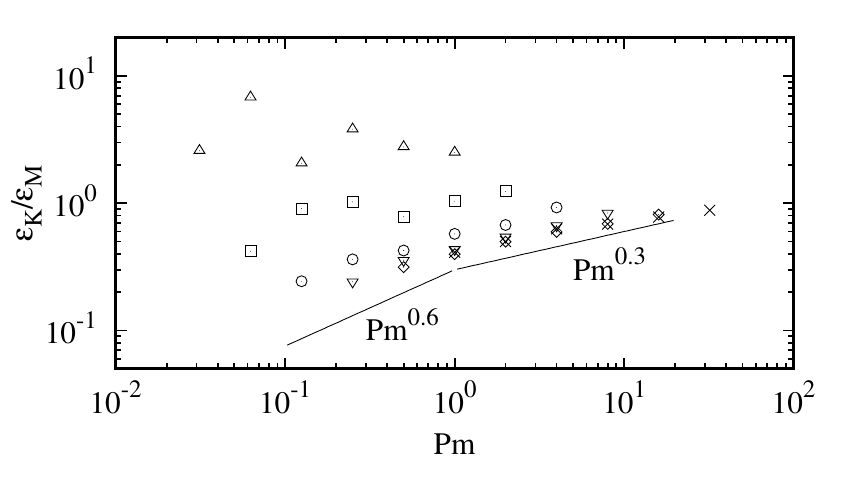}
  \caption{\label{fig:pm_e}Time-averaged kinetic-to-magnetic dissipation rate ratios
  grouped according to resistivity, $\eta$. }
\end{figure}
\begin{figure}
  \includegraphics[width=\linewidth]{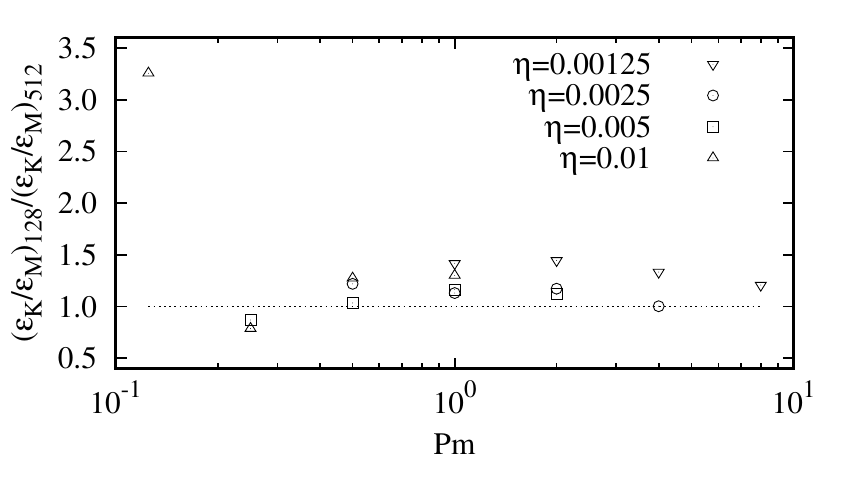}
  \caption{\label{fig:res}
  Comparison of the time-averaged kinetic-to-magnetic dissipation rate in simulations
  on a $128^3$ lattice ($\epk/\epm)_{128}$ and on a $512^3$ lattice
  $(\epk/\epm)_{512}$ with otherwise identical initial conditions.}
\end{figure}
Figure \ref{fig:pm_e} shows the kinetic-to-magnetic
dissipation ratios for our dataset.
Our $\text{Pm}>1$ data collapse onto the same line as Rm
increases, implying asymptotic independence from Rm when $\text{Pm}>1$.
The scalings for nonhelical MHD with $\text{Pm}<1$ and $\text{Pm}>1$ 
that were proposed in Ref.~\cite{Brandenburg2014} have been indicated.
Since for $\text{Pm}<1$ the kinetic dissipation scale was not properly resolved
in the simulations reported in Ref.~\cite{Brandenburg2014}, 
it is probable that the measurement of $\epk$ 
was affected,
and similarly $\epm$ when $\text{Pm}>1$,
so the steepness of the scaling of $\epk/\epm$
with Pm appears exaggerated for both $\text{Pm}<1$ and $\text{Pm}>1$
compared to our results.

The total dissipation rate was controlled by the large-scale energy
injection and is approximately constant across all of our simulations.
In our mechanically-forced simulations $\varepsilon_M$
is necessarily equal to the average net kinetic-to-magnetic energy
transfer rate, so the ratio $\varepsilon_K/\varepsilon_M$ can be
used as a measure of the efficiency of dynamo action.
Smaller values mean more energy is being
transferred to and dissipated via the magnetic field. The collapse of
our data onto one line as Rm increases in Fig.~\ref{fig:pm_e} 
shows that there is a
maximum dynamo efficiency which is curtailed as the magnetic Prandtl
number increases; that is, although a magnetic field is more easily
sustained at large values of Pm, it receives relatively less energy
transfer from the velocity field. 
This is consistent with other work from a very different
direction \cite{Schekochihin2004,Ponty2005} but within the same Pm
range, that also supports a diminishing of the dynamo.
At small values of Pm, $\epm$ may far exceed $\epk$, meaning
that if the kinetic-to-magnetic transfer rate is not able to match $\epm$, any magnetic field will eventually
dissipate fully.
This line onto which the data collapses has an inflexion point
about Pm=1, however, the equivalent line when plotting $\varepsilon_M/\varepsilon_T$
$(\varepsilon_T=\varepsilon_K + \varepsilon_M)$ as a function of Pm 
shows no such inflexion.
This serves as one explanation for the origin of the scaling behavior
of the dissipation ratio.


To illustrate the importance of resolution
we repeated on a $128^3$ lattice our simulations which had been done on a $512^3$ lattice;
see Fig.~\ref{fig:res}.
The low-resolution simulations
miscalculated the dissipation ratios by up to $40\%$,
with the biggest discrepancies mostly occuring at high Rm.
Additionally, for $\text{Pm}=1/8$, where
dynamo action was not sustainable, the low-resolution
dissipation ratio was more than 3 times the high-resolution ratio.

Analyses of triad interactions and shell-to-shell energy transfers show that
energy is transferred from the velocity field at the forcing scale to the
magnetic and velocity fields at all scales 
in a way that depends on the separation between the giving and receiving scales
and the energy contained in the involved scales,
amongst other things \cite{Waleffe1992,Linkmann2016a,Linkmann2017a,Mininni2010,Alexakis2005b,Kumar2015}. 
Therefore it is reasonable to expect a consistent scaling of $\epk/\epm$
with Pm that is not affected by whether 
$\text{Pm}<1$ or $\text{Pm}>1$, as we 
see in Fig.~\ref{fig:pm_e}.
Furthermore, when the velocity field is turbulent over a larger
range of scales than the magnetic field, i.e.~$k_\nu > k_\eta$ and $\text{Pm}<1$, 
then for a given Rm there should be a
corresponding value of Pm below which more energy will be transferred to the dissipative
part of the magnetic field, $k>k_\eta$,
 than to $k<k_\eta$.
It thus seems natural that
 the magnetic field would become unsustainable
at some critical value of Pm, as put forward in Ref.~\cite{Schekochihin2004}.
The coupling between the small-scale velocity field
and the large-scale magnetic field may be key to tipping the balance in favour of
sustainable dynamo action for small values of Pm \cite{Boldyrev2004}.
Indeed, this explains why the $\text{Pm}=1/8$ result in Fig.~\ref{fig:res}
was so large: dynamo action in the low-resolution simulation was suppressed.

\begin{figure}
  \includegraphics[width=\linewidth]{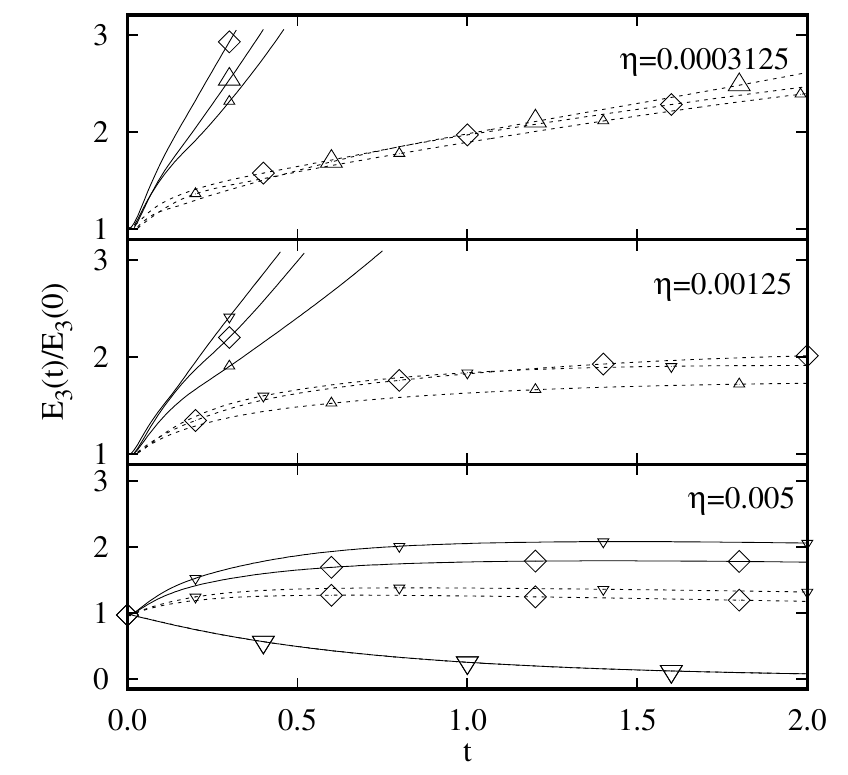}
  \caption{\label{fig:IC} $E_3(t)$ normalised by $E_3(0)$ for nonhelical runs
            (dashed lines) and helical
            runs (solid lines). Lines with diamond points correspond to $\text{Pr}_{\text{M}}=1$,
            upwards-pointing small and large
            triangles to $\text{Pm}=4$ and 16, and downward-pointing
            small and large triangles to
            $\text{Pm}=1/4$ and $\text{Pm}=1/16.$}
\end{figure}

\subsection{Reverse spectral transfer}

In Fig.~\ref{fig:mag_spec} the high-Rm data have more of
a build-up of magnetic energy in the largest scales than the lower-Rm data.
Inspired by this, we move on to examining the effect of Rm and Pm on RST
by comparing simulations of decaying MHD turbulence
with initially fully helical or nonhelical magnetic fields.
We performed 9 pairs of simulations
covering the range $1/16\leq\text{Pm}\leq16$
in multiples of 4, with the extreme values of $\nu$ and $\eta$ 
being 0.005 and 0.0003125 (see Fig.~\ref{fig:params}).
To facilitate RST, we set the peak of the initial energy spectra to $k_0=40$.

We define the energy in the first 3 wavenumbers of the magnetic field 
as $E_3(t)=\int_0^3 E_M(k,t) dk$.
Since the system is not subject to an external force,
then if $E_3(t)$ is constant or increasing,
energy must be coming from smaller length scales.
We measured $E_3(t)$ until the simulation entered a power law decay of total energy 
and plotted the results in Fig.~\ref{fig:IC}.
We found that increasing Pm by increasing Rm enhances the growth rate of RST,
with a stronger effect than increasing Pm by decreasing Re.
This indicates that RST should be possible as long as there is adequate separation of
$k_1$, $k_0$ and $k_\eta$, where $k_1=1$ is the largest wavenumber in the system
and $k_\nu $ is close to the value of $k_\eta$ or greater.
In general the high-Rm simulations (top plot in Fig.~\ref{fig:IC})
had the most RST.
RST was absent at $\text{Pm}=1/16$ but present at $\text{Pm}=1/4$ for high enough Rm.
As far as we are aware, nonhelical RST for $\text{Pm}<1$ has not been seen in previous DNS,
and may be of interest in
geophysical applications \cite{Mininni2006}.

\section{Conclusions}

The fully-resolved simulations developed in this paper
are a definitive dataset, improving confidence 
on the scaling and energy transfer properties of MHD in the near
couple decade region of magnetic Prandtl number around unity.  
We have shown that many results rely on reaching a critical Rm before we
find asymptotic dependence on Pm.
Furthermore, underresolved simulations may exaggerate the scaling of
properties such as
$\epk/\epm$ by failing to account for 
all of the dissipative dynamics.
Although our simulations feature simple geometry and do not take into account e.g.~rotation,
approaching complex physical problems from this angle may still have merit.
In black hole accretion disks, luminosity is influenced by the dissipation ratios
and DNS measurements could be a useful calibration tool.
We reiterate that fully-resolved simulations
such as ours are vital for
accurately producing dynamo action
and other effects incurred by nonunity Pm.\\

\begin{acknowledgments}
This work used the ARCHER UK National Supercomputing Service \cite{ARCHER},
made available through the Edinburgh Compute
and Data Facility (ECDF, \cite{eddie}) and the Director's Time. 
AB acknowledges funding from the Science and Technology Facilities Council.
MEM and RDJGH were supported by the Engineering and Physical Sciences Research Council (EP/M506515/1).
The data and simulation details are publicly available online \cite{datashare2}. 
\end{acknowledgments}

\bibliographystyle{apsrev4-1}

\end{document}